# Yb-doped all fiber picosecond laser based on grade-index multimode fiber with microcavity

Zhipeng Dong, Shujie Li, Ruishan Chen, Hongxun Li, Chun Gu, Peijun Yao, Lixin Xu*

*Department of Optics and Optical Engineering, University of Science and Technology of China, Hefei 230026, China*

*Corresponding author: xulixin@ustc.edu.cn

**Abstract**:We demonstrate an all-fiber mode-locked laser at all-normal dispersion with a saturable absorber, which is based on a section of grade-index multimode fiber with inner microcavity embedded in the splice points. The absorption modulation depth of this nonlinear saturable absorber is 2.2% in the suitable bent state. The fiber laser operates at central wavelength of 1035nm with average power of ~1.26mW, pulse-duration of 17.05 ps, repetition rate of 37.098MHz. This kind of saturable absorber has superiorities of low cost, high stability and high damage threshold, which may facilitate applications in optical sensing, material processing and light detection.

## 1. Introduction

Mode-locked Yb-doped ultrashort pulse fiber laser have wide applications in industrial processing and scientific research [1-5]. In order to generate mode-locked ultrashort pulse, nonlinear saturable absorber (SA) is a critical section in fiber laser. There are various saturable absorber materials such as semiconductor saturable absorber mirrors (SESAMs) [6-7], graphene [8-11], single-walled carbon nanotubes [12, 13] and $WS_2$ [14]. Even some of these have been used in commercial lasers products, they still have deficiencies such as low damage thresholds, expensive prices, narrow working bandwidth and complex production. Although, nonlinear amplifier loop mirror (NALM) or nonlinear polarization rotation (NPR) also can be used in fiber laser as an SA, their environmental sensitivity and periodic SA curve with respect to the pulse power restrict their further development in ultrahigh-peak power mode-locked lasers.

New nonlinear dynamics in grade-index multimode fiber (GIMF) have been reported such as supercontinuum generation [15-17], spatiotemporal mode-locked [18-19], self-beam cleaning [17, 20, 21, 22], soliton molecules [18], second harmonic generation [23]. In GIMF, based on the nonlinear multimode interference (NMI), a novel saturable absorber can be developed due to its nonlinear switching property. In 2003, Nazemosadat and Mafi have theoretically proved that SMF-GIMF-SMF geometry can act as saturable absorber with more than 90% modulation depths [24]. In this case, the length of fiber must meet the specific length [25]

$$L=16n_{co}a^2/\lambda \qquad (1)$$

Where $n_{co}$ is the refractive index of the multimode fiber-core and $a$ is the radius of the multimode fiber-core. However, it is very difficult to control the length of the fiber on the hundreds micron scale in practice.

Recently, it has been experimentally verified that GIMF can act as SA based on nonlinear multimode interference mechanism in Tm-doped, Er-doped, Yb-doped mode-locked or Q-switch fiber laser [26-30]. Adopting methods proposed above, it is not essential limit on the precisely choosing the length of the GIMF. Especially, Wang *et al.* are the first use GIMF as saturable absorber to generate mode-locked pulses [28]. Yang *et al.* placed a graded-index multimode fiber with an inner micro-cavity structure to achieved saturable absorber behavior where the absorption modulation depth of the saturable absorber is 1.9% [29]. Utilizing the nonlinear multimode interference effects as an SA in fiber is an attractive alternative approach to achieve ultrashort mode-locked laser pulse because of its intrinsically high power damage threshold, all-fiber format, and low cost.

Here, we constructed a stable Yb-doped mode-locked fiber laser based on GIMF with an inner micro-cavity as an SA. The absorption modulation depth of this device is 2.2% with saturation intensity of 2.25MW/cm$^2$. This is the first time that the mode-locked laser was operating in the 1 μm spectral region based on SMF-GIMF-SMF structure with an inner micro-cavity. The Yb-doped fiber laser emits pulses with pulse-width of 12.4 ps at the central wavelength of 1035nm，and its signal-to-noise ratio is ∼60 dB.

## 2. Device fabrication and operation principle

When single-mode light is coupled to a multimode waveguide, a number of high-order mode will be excited, and transmit along the fiber in a periodic interference pattern. The mode field of fiber can quasi-reproduce at certain positions during transmission which is called as Self-imaging effect. The self-imaging effect in multimode fibers (MMFs) was first observed by Allison [31] and it has been used in many application by various groups such as bandpass filter [32], fiber-optic sensors [33]. Due to the cross-phase modulation (XPM) and self –phase modulation (SPM) effects, the self-imaging period of high-peak-power are different from that of low-peak-power. This results in the performance of saturable absorption at a specific state. According to Mafi *et al*, the transmission of the light intensity from GIMF to single mode fiber (SMF) are affected by the following four factors: the number of excited modes $M$ in GIMF, the mode-field overlapping areas ratio between fundamental mode of the SMF and the LG$_{00}$ mode in GIMF $\eta$, the length of the GIMF $L$, and the total optical power $P$ [24]. Adopting micro-cavity can excite more high-order modes in fiber. In addition, bending the GIMF, the mode field distribution of GIMF will change within a certain range, which results in influence on the coupling efficiency from GIMF to SMF. When appropriate bending state is reached, optical signal of high-peak-power can transit from GIMF to the SMF at low loss, while low-peak-power optical signal will suffer a large loss. Obviously, it act as an SA. In this case, the length of the fiber is not strictly limited, which means the device can be used in mode-locked fiber laser with a flexible means.

In this paper, we fabricated a micro-lens in a section of cleaved GIMF (62.5/125, Corning) by hydrofluoric acid with the concentration of 40% for 5 minutes [29]. Because of the different corrosion rate between fiber core and cladding, a micro-lens was formed with diameter about 42.94um. Then the etched GIMF is directly spliced with SMF by a fusion splicer (Fujikura 60s). The microscope image of etched optical fiber is presented in figures 1(a). In our experiment, the length of GIMF is estimated to be 14cm. Figure. 1 (b) shows the microscope image of the inner-cavity by fusion splicer. Figure. 1 (c) shows the schematic diagram of the SA constructed.

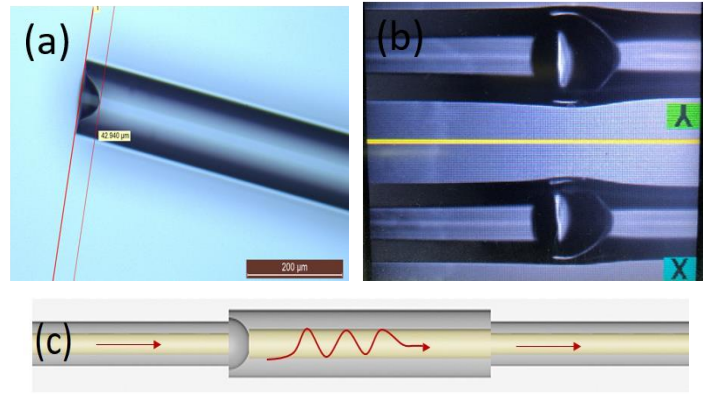

Fig. 1. (a) The microscope images of device sample; (b) Schematic diagram of the device fabrication; (c) The schematic diagram of SMF-GIMF-SMF structure.

We obtain the transmittance of this device by a mode-locked fiber laser, which operates at 1030nm with 3dB bandwidth of 15nm with a repetition rate of 7.83MHz. Figure. 2 shows the effective transmission dependence on the pulse intensity. For simplification, the transmission curve of this device can be described as [34]

$$T(I)=1- \alpha \times \exp(-I/I_{sat}) - \alpha_{ns} \tag{1}$$

where $T$ is the transmittance, $\alpha$ is the modulation depth, $I$ is the input light intensity, $I_{sat}$ is the saturation intensity, and $\alpha_{ns}$ is the nonsaturable loss. According to the above formula, the experimental data can be fitted into a curve. The absorption modulation depth is measured as 2.2% and the value of saturation fluence is ∼2.25MW/cm$^2$. Table I illustrates the optical properties of other SA materials. The comparison shows that the modulation depth are about the same as that of other SAs [8, 12, 14,], and the saturable intensity is relatively lower.

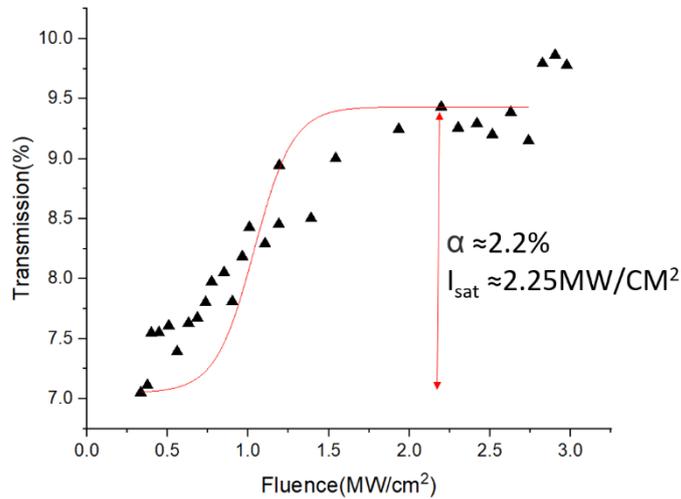

Fig. 2. Measured transmission curves with different bending degrees of SA.

TABLE I
OPTICAL PROPERTIES OF DIFFERENT SAs

| SA | MODULATION DEPTH | SATURATION INTENSITY (MW/CM2) |
|---|---|---|
| Graphene[8] | 2.93% | 53.25 |
| single-walled carbon nanotubes[12] | 4.3% | 34 |
| WS2[14] | 1.2% | 25 |
| SIMF-GIMF device | 3.16% | 1.34 |
| Bi2Te3 [34] | 27% | 0.058 |

## 3. Experimental results and discussion

The configuration of mode-locked fiber laser structure is presented in Fig. 3. The laser cavity consists of a 980nm laser diode pump with maximum pump power 450mW, a 980/1030nm wavelength division multiplexing (WDM), a section of 50cm highly ytterbium-doped fiber (LIEKKI YB1200). An intra-cavity polarization-independent isolator is employed to ensure the unidirectional operation of the ring cavity. A polarization controller (PC) is used to fine-tune the state of polarization in the cavity. The length of GIMF (corning 62.5/125) with an inner cavity is about 14cm, which acts as an SA in fiber laser. Apart from ytterbium doped fiber and GIMF, other fibers in ring cavity are standard single-mode fiber. The total cavity length is 5.4m. The 1% energy of the mode-locked pulses is coupled out by using a 99:1 output coupler. In order to monitor optical spectrum and the time evolution of the laser pulse simultaneously, an 80:20 coupler is connected to spectrometers and oscilloscopes respectively. The pulse spectrum is measured by an optical spectrum analyzer (OSA, ANDO AQ6317B), and temporal properties are measured by a 4 GHz digital oscilloscope (LeCroy Wave Runner 640Zi ) with a 3GHz photodetector, and using 2Hz–2GHz radio-frequency (RF) spectrum analyzer to obtain pulse train of signal-to-noise. The pulse width was obtained by a commercial autocorrelator (APE pulseCheck 600).

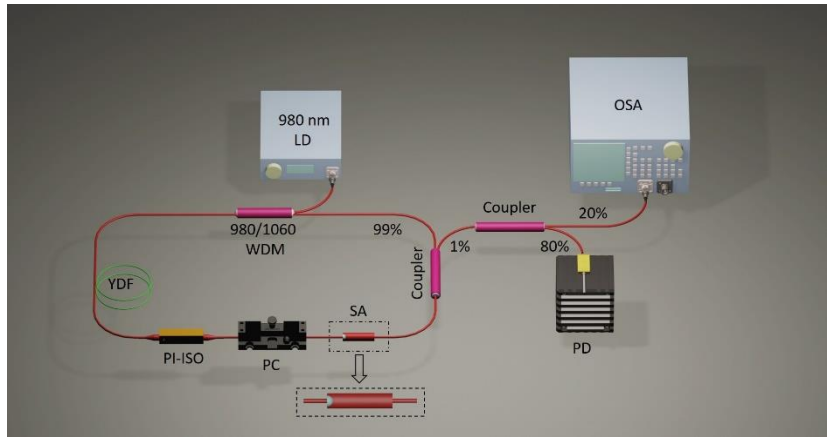

Figure. 3. Experimental setup of the mode-locked fiber laser.

In this fiber laser, the mode-locked pulses are achieved when the pump power reached 140mW. Fig. 4. shows the characteristics of the fiber laser output power. At the maximum pump power 430mW, the fiber laser emits 1.26mW average power pulses train. As illustrated in Fig. 5 (a), the spectrum of the mode-locked fiber laser is centered at 1037 nm with 3dB bandwidth of 4 nm, the smooth spectrum suggests that the fiber laser emits a stable pulses train. In order to confirm the operation stability, RF spectrum is measured with a 10 Hz resolution bandwidth and 500 KHz scan range, as illustrated in Fig. 5 (b). The result indicates the high stability of mode-locked pulses. The inset of Fig. 5. (b) shows the frequency range from 0 Hz to 1 GHz without any sidebands generation, which also confirm the high stability of the mode-locked pulses again. Figure. 5 (c) shows that the period of laser output pulse train was 26.96ns, which was corresponding to cavity round-trip time and verified the fiber laser was operating in the single pulse state. The inset of Fig. 5 (c) is trace diagram of oscilloscope with 0.4 $\mu s$ time scan range. A measured autocorrelation trace is shown in Fig. 5 (d). The trace was fitted to a Gaussian shape and the pulse duration was estimated to be ∼17.6 ps.

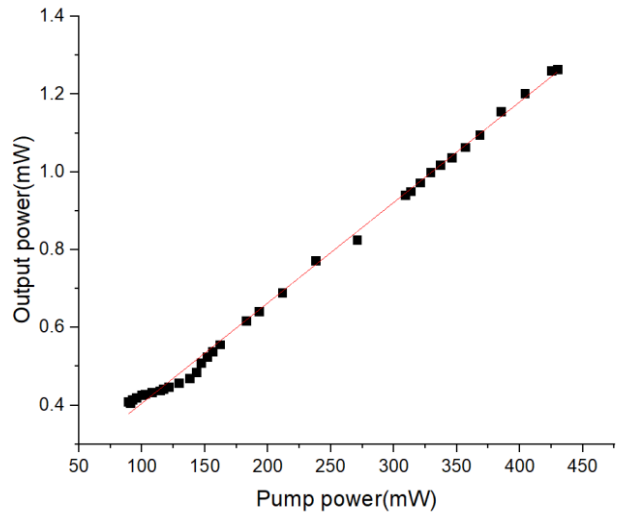

Fig. 4. Output characteristics as a function of the pump power.

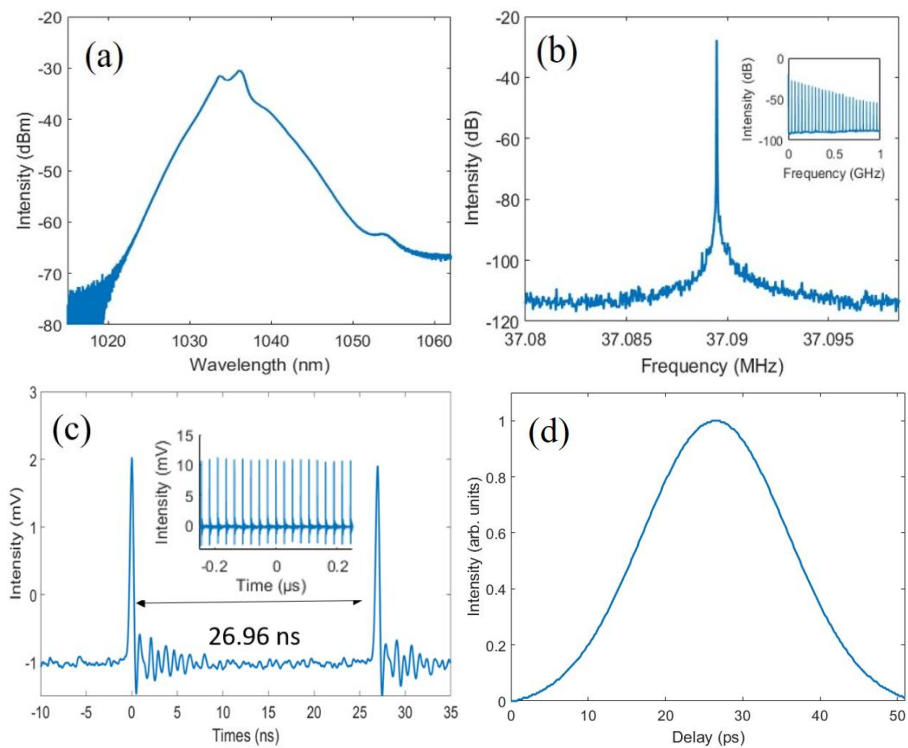

Fig. 5. Mode-locked pulse measurements: (a) Laser spectrum. (b) RF spectrum with repetition rate of 37.098 MHz, span of 500 KHz, and resolution bandwidth of 10 Hz. Inset: RF spectrum over a frequency range of 0-1 GHz. (c) pulses train. Inset: oscilloscope trace with a time duration of 0.4 μs (d) Autocorrelation trace of output pulses obtained by Oscilloscope.

To verify the mode-locked operation was purely contributed by this SA structure, the all-fiber SA was purposely removed from the laser cavity. Then, no matter how we adjusted the PC or increasing the pump power, the fiber laser could not output stable pulses train.

**4. Conclusion**

In summary, we have developed a stable mode-locked Yb-doped fiber laser based on grade-index multimode fiber with an inner microcavity. By adopting this structure, a simple and stable SA is achieved. The ultrashort pulses are obtained with pulse width of 17.05 ps, and spectral center wavelength is 1037nm. At maximum pump power of 430mW, the fiber laser has average power of 1.26 mW at the 37.098MHz repetition rate. We believe that this type fiber laser of low cost, high stability and high damage threshold provide the potential applications in precision measurements, nonlinear research, industrial processing.